\documentclass[12pt]{article}
\usepackage[dvips]{graphicx}
\def\doublespace{\lineskip      .25 ex\baselineskip 3.0
ex\lineskiplimit 0 ex\parskip 1.0 ex plus.50 ex minus .25 ex}%

\begin{document}
\doublespace

\title{The Schwarzschild's Braneworld Solution}
\author{ J.
Ovalle\footnote{jovalle@usb.ve}
 \\
\vspace*{.25cm}\\
 Departamento de F\'{\i}sica, Universidad Sim\'on Bol\'{\i}var \\
Caracas, Venezuela\\
 }
\date{}
\maketitle
\begin{abstract}
In the context of the Randall-Sundrum braneworld, the minimal geometric deformation approach, which has been
successfully used to generate exact interior solutions to
Einstein's field equations for static braneworld
stars with local and non-local bulk terms, is used to obtain the braneworld version of the Schwarzschild's interior solution. Using this new solution, the behaviour of the Weyl functions is elucidated in terms of the compactness for different stellar distributions.

\end{abstract}
\newpage
\section{Introduction}

The study on braneworld consequences in general relativity has been extensively studied during the last years \cite{maartRev2004}, \cite{maartRev2010}. 
However many unknowns remain unanswered, within which the non-closure of the braneworld equations is undoubtedly one of the most important issues. 
To overcome this problem a better understanding of the bulk geometry and proper boundary conditions is required. Despite the fact that there is not 
a definitely answer to these issues, there is an approach that allows to generate the braneworld version of every known general relativistic solution, 
the so called {\it minimal geometric deformation approach} (the {\it minimal anisotropic consequence} in the terminology of Ref.\cite{jovalle2009}). This route has been successfully used in the astrophysics context \cite{jovalle07}, 
allowing elucidate some aspects of the roll played by the Weyl fluid inside a stellar distribution \cite{jovalle207}.

On the other hand, it is well known that the Schwarzschild's interior metric is the only stable solution for a bounded distribution \cite{weinberg} which fits smoothly with the Schwarzschild exterior metric. Therefore the study of this interior solution in the braneworld context represents a scenario of great interest. Indeed, in the pioneer work of Germani and Maartens
\cite{germ} the braneworld generalization of the Schwarzschild's interior solution was reported, but only high energy corrections were considered, leaving out the analysis of the effects of Weyl functions inside uniform distributions. Unfortunately, so far there has not been possible to build a consistent braneworld version of this interior solution when a Weyl fluid is considered along with high energy corrections, mainly because the non-close and non locality problems complicate tremendously the braneworld scenario. The existence of such a solution would be very useful to elucidate the role played by Weyl fluids inside uniform stellar distributions.

In the context of uniform stars in the braneworld, it is also important to note that the collapse of
a homogeneous star leads to a non-static exterior solution
\cite{bgm}-\cite{kp2004}. Indeed, Govender and Dadhich
\cite{mgnd} proved that a collapsing sphere on the brane radiates,
and they shown that the exterior for this radiative sphere can be
described by a Vaidya metric that envelops the collapsing region.
However, in this work only static uniform distributions will be considered, leaving aside 
for now the analysis of gravitational collapse.

In this paper a consistent version of the Schwarzschild's interior metric is build in the context of the braneworld, where local bulk
terms (high energy corrections) and non-local bulk terms (bulk
Weyl curvature contributions) are considered. Using the minimal geometric deformation approach,
 all problems associated with the searching of braneworld solutions are overcome. 
The solution found is used to describe in detail the behaviour of both Weyl function, namely 
the scalar ${\cal U}$ and the anisotropy ${\cal P}$, inside the stellar distribution, showing 
thus that in general both of them are proportional to the compactness of the stellar distribution. 
It is shown that the behaviour of the Weyl functions inside the stellar distribution may be easily 
interpreted in terms of the geometric deformation undergone by the radial metric component due to 
five dimensional effects. Also, it is shown that it is possible to obtain a model where the pressure is 
increased by five dimensional effects, which represents a different result from other braneworld 
solution \cite{germ}, where non-local terms were considered, thus showing that the bulk Weyl 
curvature contributions have important consequences.

\section{Field equations and the minimal geometric deformation approach.}

 The effective Einstein's field equation in the brane can be written as a modification of the standard field equation through an energy-momentum tensor carrying bulk effects onto the brane:
\begin{equation}\label{tot}
T_{\mu\nu}\rightarrow T_{\mu\nu}^{\;\;T}
=T_{\mu\nu}+\frac{6}{\sigma}S_{\mu\nu}+\frac{1}{8\pi}{\cal
E}_{\mu\nu},
\end{equation}
here $\sigma$ is the brane tension, with  $S_{\mu\nu}$ and
$\cal{E}_{\mu\nu}$ the high-energy and  non-local corrections respectively.

Using the line element in Schwarzschild-like coordinates
\begin{equation}
\label{metric}ds^2=e^{\nu(r)} dt^2-e^{\lambda(r)} dr^2-r^2\left( d\theta
^2+\sin {}^2\theta d\phi ^2\right)
\end{equation}
in the case of a spherically symmetric and static distribution having Weyl stresses in the interior, the effective equations can be written as
\begin{equation}
\label{usual} e^{-\lambda}=1-\frac{8\pi}{r}\int_0^r r^2\left[\rho
+\frac{1}{\sigma}\left(\frac{\rho^2}{2}+\frac{6}{k^4}\cal{U}\right)\right]dr,
\end{equation}
\begin{equation}
\label{pp}\frac{8\pi}{k^4}\frac{{\cal
P}}{\sigma}=\frac{1}{6}\left(G_1^1-G_2^2\right),
\end{equation}
\begin{equation}
\label{uu}\frac{6}{k^4}\frac{{\cal
U}}{\sigma}=-\frac{3}{\sigma}\left(\frac{\rho^2}{2}+\rho
p\right)+\frac{1}{8\pi}\left(2G_2^2+G_1^1\right)-3p
\end{equation}
\begin{equation}
\label{con1}p_{1}=-\frac{\nu_1}{2}(\rho+p),
\end{equation}
with
\begin{equation}
\label{g11} G_1^1=-\frac 1{r^2}+e^{-\lambda }\left( \frac
1{r^2}+\frac{\nu _1}r\right),
\end{equation}
\begin{equation}
\label{g22} G_2^2=\frac 14e^{-\lambda }\left[ 2\nu _{11}+\nu
_1^2-\lambda _1\nu _1+2 \frac{\left( \nu _1-\lambda _1\right)
}r\right].
\end{equation}
where $f_1\equiv df/dr$ and $k^2=8{\pi}$. The general relativity
is regained when $\sigma^{-1}\rightarrow 0$ and (\ref{con1})
becomes a lineal combination of (\ref{usual})-(\ref{uu}).

The Eqs.(\ref{usual})-(\ref{con1}) represent an indefinite system of
equations in the brane, a problem that essentially is represented by the fact that there is only one equation, that is, the conservation equation (\ref{con1}), to find three unknown functions
$\{p(r), \rho(r), \nu(r)\}$. Hence to close the system in the brane we have to consider some restrictions. However this is an open problem for which the solution requires  more information of the bulk geometry and a
better understanding of how our 4D spacetime is embedded in the
bulk. Despite the above, it is possible to generate the braneworld version of every general relativistic solution through the {\it minimal geometric deformation approach} (or minimal anisotropic consequence), as explained briefly next (all details are shown in \cite{jovalle2009}).

The first step is to pick up a known solution $\{p(r), \rho(r), \nu(r)\}$ to Eq.(\ref{con1}). Thus the problem is reduced to solve the integral differential equation for the geometric function $\lambda(r)$ shown in Eq.(\ref{usual}). To accomplish this the following solution is proposed
\begin{eqnarray}
\label{expect}  e^{-\lambda}=1-\frac{8\pi}{r}\int_0^r r^2\rho
dr+(Bulk\;\;effects),
\end{eqnarray}
that is
\begin{eqnarray}
\label{expect}  e^{-\lambda}=\underbrace{1-\frac{8\pi}{r}\int_0^r r^2\rho
dr}_{ \mu}\; +\; Geometric\;\;Deformation.
\end{eqnarray}
The unknown {\it geometric deformation} in (\ref{expect}) should have two sources: extrinsic curvature and five dimensional Weyl curvature, hence it can be written as a generic $f$ function
\begin{eqnarray}
\label{expectg}  e^{-\lambda}=\mu+f
\end{eqnarray}
which at the end will have the form
\begin{equation}
\label{deform}
f= \frac{1}{\sigma}(\;high\;energy\;terms)\; + \;non\;local\;terms.
\end{equation}
Finally the solution to the integral differential equation (\ref{usual}) is found to be
\begin{equation}
\label{edlrwss} e^{-\lambda}={1-\frac{8\pi}{r}\int_0^r r^2\rho
dr}+\underbrace{e^{-I}\int_0^r\frac{e^I}{(\frac{\nu_1}{2}+\frac{2}{r})}\left[H(p,\rho,\nu)+\frac{8\pi
}{\sigma}\left(\rho^2+3\rho p\right)\right]dr}_{f},
\end{equation}
with
\begin{eqnarray}
\label{I} I\equiv
\int\frac{(\nu_{11}+\frac{\nu_1^2}{2}+\frac{2\nu_1}{r}+\frac{2}{r^2})}{(\frac{\nu_1}{2}+\frac{2}{r})}dr
\end{eqnarray}
and the function $H(p,\rho,\nu)$ defined as
\begin{equation}
\label{H} H(p,\rho,\nu)\equiv
\left[\mu_1(\frac{\nu_1}{2}+\frac{1}{r})+\mu(\nu_{11}+\frac{\nu_1^2}{2}+\frac{2\nu_1}{r}+\frac{1}{r^2})-\frac{1}{r^2}\right]-8\pi
3p.
\end{equation}
In order to recover general relativity, the following condition
must be satisfied
\begin{equation}
\label{constraintf} lim_{{\sigma}^{-1}\rightarrow\;\;
0}\;\;\int_0^r\frac{e^I}{(\frac{\nu_1}{2}+\frac{2}{r})}H(p,\rho,\nu)dr
=0,
\end{equation}
but this constraint is automatically satisfied by every general relativistic solution, since each of these solutions satisfy the constraint
\begin{equation}
\label{constraint2} H(p,\rho,\nu)=0.
\end{equation}
On the other hand, when the constraint (\ref{constraint2}) hold, the anisotropy induced onto the brane due to the geometric deformation undergone by $\lambda(r)$ may be written as
\begin{equation}
\label{ppf3}
\frac{48\pi}{k^4}\frac{{\cal P}}{\sigma}=
\underbrace{(G_1^1-G_2^2)\mid_{\frac{1}{\sigma}=0}}_{=\,0}+f^{*}(\frac{1}{r^2}+\frac{\nu_1}{r})-\frac{1}{4}f^{*}(2\nu_{11}+\nu_{1}^2+2\frac{\nu_1}{r})-\frac{1}{4}f^{*}_1(\nu_1+\frac{2}{r}),
\end{equation}
where
\begin{equation}
\label{mindef}
f^{*}=\frac{1}{\sigma}(\;high\;energy\;terms)\; + \underbrace{\;non\;local\;terms}_{=\,0}
\end{equation}
is the {\it minimal geometric deformation}, whose explicit form may be seen by (\ref{edlrwss}) as following
\begin{equation}
\label{fsolutionmin}
f^{*}=\frac{8\pi
}{\sigma}e^{-I}\int_0^r\frac{e^I}{(\frac{\nu_1}{2}+\frac{2}{r})}\left(\rho^2+3\rho p\right)dr.
\end{equation}
The expression (\ref{fsolutionmin}) represents a minimal geometric deformation in the sense that all sources of the geometric deformation $f$ have been removed except those produced by the density and pressure, which are always present in a stellar distribution\footnote{An even minimal deformation is obtained for a dust cloud, where $p=0.$}. It is clear that this minimal deformation will produce a minimal anisotropy onto the brane.

\section{The Schwarzschild's solution in the braneworld}

Let us construct the braneworld version of Schwarzschild's
solution using the approach describe in the previous section. The general Schwarzschild's interior solution in general relativity is given by
\begin{equation}\label{schw00}
e^{\nu}=\left(A-B\sqrt{1-r^2/C^2}\right)^2,
\end{equation}
\begin{equation}\label{schw11}
e^{-\lambda_S}=1 - \frac{r^2}{C^2},
\end{equation}
\begin{equation}\label{schwdensity}
\rho =\frac{3}{8\pi C^2 },
\end{equation}
and
\begin{equation}
\label{schwpressure} p(r)=\frac{\rho}{3}\left[\frac{{3B\sqrt{1 -
\frac{r^2}{C^2}}} - A}{A-B{\sqrt{1 -
\frac{r^2}{C^2}}} }\right]
\end{equation}
where $A$, $B$ \footnote{The value $B=\frac{1}{2}$ is not necessary
true for braneworld stars, as there are many possible exterior
solutions.} and $C$ are constants to be determined by matching
conditions. Using the condition \footnote{This condition can be dropped for braneworld stars
\cite{deru}, \cite{gergely2007}} $p=0$ at the stellar surface $r=R$, the following relationship among $A$, $B$ and $C$ is found
\begin{equation}
\label{ABC}
A=3B\sqrt{1 -
\frac{R^2}{C^2}},
\end{equation}
yielding
\begin{equation}\label{schwnu}
e^{\nu}=B^2\, {\left(
          3\, {\sqrt{1 - \frac{R^2}{C^2}}}-{\sqrt{1 - \frac{r^2}{C^2}}}  \right) }^2,
\end{equation}
and
\begin{equation}
\label{schwpressure} p(r)=\rho\left[\frac{{\sqrt{1 -
\frac{r^2}{C^2}}} - {\sqrt{1 - \ \frac{R^2}{C^2}}}}{3\, {\sqrt{1 -
\frac{R^2}{C^2}}}-{\sqrt{1 - \frac{r^2}{C^2}}} }\right]
\end{equation}

The expressions (\ref{schwdensity}), (\ref{schwnu}) and
(\ref{schwpressure}) automatically satisfy the constraint
(\ref{constraint2}), and the bulk effects on them are found through the
matching conditions. On the other hand, the braneworld version for
the radial metric component (\ref{schw11}) is obtained using
(\ref{schwdensity}), (\ref{schwnu}) and (\ref{schwpressure}) in
(\ref{edlrwss}), leading to
\begin{equation}\label{reglambda}
e^{-\lambda(r)}=1-\frac{2\tilde{m}(r)}{r},
\end{equation}
where the interior mass function $\tilde{m}$ is given by
\begin{eqnarray}\label{regularmass}
\tilde{m}(r)&=&m(r)-\frac{1}{\sigma}\frac{9\, r}{16 \pi\,C^4
}g(r),
\end{eqnarray}
with $m(r)$ being the general relativity interior mass function,
given by the standard form
\begin{equation}
\label{regularmass2} m(r)=\int_0^r 4\pi r^2{\rho}dr=\frac{r^3}{2
C^2},
\end{equation}
and
\begin{eqnarray}
\label{gr} g(r)&=&e^{-I(r)}\int_0^r \frac{e^{I(r')}\,
    r'\, \left(1- r'^2/C^2\ \right) }{3\frac{r^2}{C^2}-2+2\gamma\sqrt{1-\frac{r^2}{C^2}}}\,
            dr',
\end{eqnarray}
where 
\begin{equation}
\label{gamma}
\gamma\equiv\,3\sqrt{1-\frac{R^2}{C^2}}
\end{equation}
and $I$ is given by Eq. (\ref{I}), which in general may be written as
\begin{equation}
\label{Igeneral}
I=\nu+ln\left(\frac{\nu_1}{2}+\frac{2}{r}\right)^2+6\,\int\,\frac{dr}{r^2\left(\frac{\nu_1}{2}+\frac{2}{r}\right)}.
\end{equation}
In this case
\begin{equation}
 e^{6\int\,\frac{dr}{r^2\left(\frac{\nu_1}{2}+\frac{2}{r}\right)}}=
\left[\frac{\gamma-3\sqrt{1-\frac{r^2}{C^2}}+\sqrt{3+\gamma^2}}{-\gamma+3\sqrt{1-\frac{r^2}{C^2}}+\sqrt{3+\gamma^2}}\right]
^{\frac{-\frac{\gamma}{2}}{\sqrt{3+\gamma^2}}}\frac{\frac{r^3}{C^3}}{\sqrt{3\frac{r^2}{C^2}-2+2\gamma\sqrt{1-\frac{r^2}{C^2}}}}
\end{equation}

The total general relativity mass is obtained through
(\ref{regularmass2}), leading to
\begin{equation}
\label{regtotmass} M\equiv m(r)\mid_{r=R}\,=\frac{R^3}{2 C^2}.
\end{equation}

Using (\ref{pp}) and (\ref{uu}) the interior Weyl functions are
written as
\begin{eqnarray}
\label{regP} {\cal P}(r)&=& \frac{1}{6C^4r^2\left(1-\frac{r^2}{C^2}\right)\sqrt{1-\frac{r^2}{C^2}}\left(
\sqrt{1-\frac{r^2}{C^2}}-\gamma\right)\left(3\frac{r^2}{C^2}-2+2\gamma\sqrt{1-\frac{r^2}{C^2}}\right)^2
   }
\nonumber \\ && \left\{9r^2( 1-\frac{r^2}{C^2})^{2}\, 
\left[ 6\frac{r^4}{C^4}-(7+2\gamma^2-7\gamma\sqrt{1-\frac{r^2}{C^2}})\frac{r^2}{C^2}+2+2\gamma^2-4\gamma\sqrt{1-\frac{r^2}{C^2}}\right]\right.
\nonumber \\ &&
             +18\,g(r)\left[36\frac{r^8}{C^8}-3\left(31+11\gamma^2-20\gamma\sqrt{1-\frac{r^2}{C^2}}\right)\frac{r^6}{C^6} \right.
\nonumber \\ &&           
+\left(88+82\gamma^2-121\gamma\sqrt{1-\frac{r^2}{C^2}}-6\gamma^3\sqrt{1-\frac{r^2}{C^2}}\right)\frac{r^4}{C^4}
\nonumber \\ &&
-\left(37+67\gamma^2-4\gamma(20+3\gamma^2)\sqrt{1-\frac{r^2}{C^2}}\right)\frac{r^2}{C^2}
\nonumber \\ &&           
                \left.\left.+6\left(1+3\gamma^2-\gamma(3+\gamma^2)\sqrt{1-\frac{r^2}{C^2}}\right)\right]
              \right\}
\end{eqnarray}
\newpage
and
\begin{eqnarray}\label{regU}
{\cal U}(r) &=& \frac{1}{12C^{4}\left(1-\frac{r^2}{C^2}\right)\sqrt{1-\frac{r^2}{C^2}}\left(
\sqrt{1-\frac{r^2}{C^2}}-\gamma\right)\left(3\frac{r^2}{C^2}-2+2\gamma\sqrt{1-\frac{r^2}{C^2}}\right)^2
   }
\nonumber \\ && \left\{9 \left(1-\frac{r^2}{C^2}\right)\left[-3\frac{r^6}{C^6}+\left(11-8\gamma^2
+\gamma\sqrt{1-\frac{r^2}{C^2}}\right)\frac{r^4}{C^4}\right.\right.
\nonumber \\ &&
\left.-4(\gamma^2-1)\left(\gamma\sqrt{1-\frac{r^2}{C^2}}
-3\right)\frac{r^2}{C^2}+4(\gamma^2-1)\left(\gamma\sqrt{1-\frac{r^2}{C^2}}
-1\right)\right]
\nonumber \\ &&
        +36\frac{g(r)}{C^2}(5-3\frac{r^2}{C^2})\left[-3\frac{r^4}{C^4}+\left(5+2\gamma^2-5\gamma\sqrt{1-\frac{r^2}{C^2}}\right)
\frac{r^2}{C^2}\right.
\nonumber \\ &&
          \left.\left.
          -2(1+\gamma^2)+4\gamma\sqrt{1-\frac{r^2}{C^2}}
              \right]\right\}.
\end{eqnarray}
Even though both Weyl functions have complicated expresions, it is possible to identify some general features. For instance, when the limit $r\rightarrow\,0$ is taken we have
\begin{equation}
\label{dark0}
 {\cal U}(0)=-\frac{3}{4\,C^4}\frac{\gamma+1}{\gamma-1},
\end{equation}
hence we may see clearly a divergence at the origen when $\gamma=1$. Then using Eq. (\ref{gamma}) and Eq. (\ref{regtotmass}) it is found that this divergence is produced when
\begin{equation}\label{compact}
 \frac{M}{R}=\frac{4}{9},
\end{equation}
obtaining thus the well known general relativistic upper bound for the compactness limit of the star. Hence it may be seen by Eq. (\ref{dark0}) that  ${\cal U}(0)$ is always negative, otherwise $\gamma\,<\,1$ would mean $M/R>4/9$, which is not allowed by general relativity. Likewise, since that $M/R<4/9$ then $C>\frac{3\sqrt{2}}{4}R$, thus $r/C\,<\,1$. Moreover, a numerical analysis on the function $g(r)$ shows that it is always positive. All this is useful to analyze the complicated expresions for both Weyl functions shown in Eq. (\ref{regP}) and Eq. (\ref{regU}). For instance, it is found that the ${\cal U}(r)$ function is allways negative, except close to the surface of stellar distributions with a compactness near the limit (\ref{compact}), as shown in figure 2 for $R=5$ and $C=5.4$ ($M/R\approx\,0.42$).

On the other hand, the anisotropy ${\cal P}(r)$ has an aparent divergence at $r=0$ due to the factor $g(r)/r^2$ in its expression. However using l'Hopital's rule it is found that 
\begin{equation}
lim_{\,r\rightarrow\,0}\frac{g(r)}{r^2}=\frac{1}{6(\gamma-1)},
\end{equation}
hence ${\cal P}(0)=0$ is obtained, in agreement with the expression given by Eq. (\ref{ppf3}). Figure 1 shows the behaviour of the anisotropy ${\cal P}$ inside the stellar distribution for different densities. It can be seen that the anisotropic stress is proportional to the density: the most compact distribution undergoes a higher anisotropic effect. This behaviour is easily explained in terms of the source of the anisotropy, which is nothing but the geometric deformation undergone by the radial metric component, explicitly shown through the solution (\ref{edlrwss}). When $H=0$ is imposed, the only source for the geometric deformation are the high energy terms presented in the expression (\ref{edlrwss}), which are quadratic terms in the density and pressure. Hence the higher the density is, the more geometric deformation will be produced, and in consequence the anisotropy induced will be higher for more compact distributions. However this situation changes near the surface, where a switched behaviour can be seen due to matching conditions. Also it can be seen that the anisotropy increases from the surface until reaches a maximum values, then decreases until ${\cal P}=0$ at $r=0$, in agreement with the explained in terms of geometric deformation, which is given explicitly by Eq. (\ref{fsolutionmin}).

On the other hand, figure 2 shows the scalar Weyl function ${\cal U}$ for the three distributions considered in figure 1. This function is more negative for more compact stellar objects except for external layers. This means that high energy terms always dominate on anisotropic terms, which are the two sources for ${\cal U}$, as can be seen through Eq. (\ref{uu}), and that this domain, which always decreases for external layers, is reduced even more for more compact distributions. Indeed, as already mentioned, the ${\cal U}$ scalar function may be positive close to the stellar surface if the object has a compactness near to the limit permitted by general relativity. Next the matching conditions are analyzed.

There are many interesting vaccum solutions in the braneworld. For instance, the solution found by Dadhich, 
Maartens, Papadopoulos and Rezania (DMPR) in Ref. \cite{dmpr} and the one obtained by Casadio, Fabbri and Mazzacurati in Ref. \cite{CFMsolution} were 
recently sucesfully considered in the solar system tests by Bohmer, De Risi, Harko and Lobo \cite{BDHL}.
In this paper the DMPR solution, given by  
\begin{equation}
\label{RegRNmet}
e^{\nu^+}=e^{-\lambda^+}=1-\frac{2\cal{M}}{r}+\frac{q}{r^2},
\end{equation}
\begin{equation}
\label{RegRNmet2} {\cal U}^+=-\frac{{\cal P}^+}{2}=\frac{4}{3}\pi
q\sigma\frac{1}{r^4},
\end{equation}
is considered as the exterior solution. Hence when the matching condition $[ds^2]_{\Sigma}=0$ at the
stellar surface $\Sigma$ is used, we have
\begin{equation}
\label{RegmatchNR1}
4\,B^2\left(1-\frac{R^2}{C^2}\right)=1-\frac{2\cal{M}}{R}+\frac{q}{R^2},
\end{equation}
\begin{eqnarray}\label{RegmatchNR2}
\frac{2\cal{M}}{R}&=&\frac{2M}{R}-\frac{1}{\sigma}\frac{9\,g(C)}{8\pi\,C^4}+\frac{q}{R^2},
\end{eqnarray}
\begin{eqnarray}
\label{qNR} \frac{q}{R^2}&=& \frac{1}{\sigma}\frac{-3\,R^2 \left(
-1 + 64\, {\pi }^2 \right)\left( 8\, C^4 - 15\, C^2\, R^2 + 7\,
R^4 \right)   }{1024\,
    C^4\, {\pi }^3 \left( 4\, C^4 - 7\, C^2\, R^2 + 3\, R^4 \right) }
\nonumber \\ &&
    +\frac{1}{\sigma}\frac{g(C)\,\left[ -768\,
        C^4{\pi }^2 + C^2\left( -5 + 896{\pi }^2 \right)
        R^2 + 3\left( 1 - 64\, {\pi }^2 \right)R^4 \right] }{1536
        {\pi }^3 \left( 4\, C^4 - 7\, C^2\, R^2 + 3\, R^4 \right) },
        \nonumber \\
\end{eqnarray}
with $g(C)\equiv g(R)$. The constants $\cal{M}$ and $q$ are given in terms of $C$ through
equations (\ref{RegmatchNR2}) and (\ref{qNR}) respectively. Using the Eq. (\ref{regtotmass}) the matching condition (\ref{RegmatchNR1}) can be written as
\begin{equation}
\label{match5}
4\,B^2\left(1-\frac{R^2}{C^2}\right)=\left(1-\frac{R^2}{C^2}\right)+\frac{1}{\sigma}\frac{9\,g(C)}{8\pi\,C^4}.
\end{equation}
There is not way to obtain any eventual bulk effect $\delta\,C$ on $C\rightarrow\,C+\delta\,C$ through the matching condition (\ref{match5}). However the expression (\ref{match5}) shows that $B$ has been modified by bulk effects as
\begin{equation}
\label{deltaB} B(\sigma)=\frac{1}{2}+\frac{1}{\sigma}\frac{1}{1-\frac{R^2}{C^2}}\,\frac{9\,g(C)}{32\,\pi\,C^4}+{\cal
O}(\sigma^{-2}).
\end{equation}
Since both the density and pressure depend only on the constant $C$, but not on $B$, there are no bulk consequences on them in this solution.

As was already mentioned, there are not five dimensional effects on the pressure or density in this solution. This is because the free parameter $C$ has no bulk correction in the form $C+\delta\,C$, as may be seen through Eq. (\ref{match5}). However there is a way to obtain the bulk correction $\delta\,C$ and in consequence the five dimensional effects on the pressure and density as explained next. Let us start from the relationship among the parameters $A$, $B$ and $C$ given through Eq. (\ref{ABC}). If $A$ is considered as a constant which is not modified by five dimensional effects, that is, $\delta\,A =0$, then necessarily there is a correction $\delta\,C$ on $C$ due to the correction $\delta\,B$ on $B$ shown in (\ref{deltaB}), which is written as
\begin{equation}
\label{deltaC}
\delta\,C=-2\frac{C^3}{R^2}\left(1-\frac{R^2}{C^2}\right)\delta\,B,
\end{equation}
thus $\delta\,C$ is finally written as
\begin{equation}
\label{deltaC2}
\delta\,C=-\frac{1}{\sigma}\frac{9}{16\,\pi\,C\,R^2}\,g(C).
\end{equation}
A numerical analysis on $g(C)$ shows that it is always positive, 
hence $\delta\,C < 0$. Figure 3 shows a qualitative comparison of the pressure 
in general relativity and braneworld when $\delta\,A=0$ for the Schwarzschild's solution. 
It may be seen that the behaviour of the pressure under five dimensional effects is 
different from other braneworld solution with uniform densities \cite{germ}, 
where internal non-local Weyl functions were not considered, thus showing that 
these non-local effects play a relevant role inside the stellar distribution.

\section{Conclusions and outlook}

In the context of the Randall-Sundrum braneworld, a consistent version of the Schwarzschild's interior 
metric was constructed, where local bulk
terms (high energy corrections) and non-local bulk terms (bulk
Weyl curvature contributions) were considered. Using the minimal geometric deformation approach, 
all problems associated with the searching of braneworld solutions are overcome. 
Indeed, both Weyl interior functions were obtained, showing that in general both of them are proportional to 
the compactness of the stellar distribution. In the case of the anisotropy stress ${\cal P}$ it was found that it 
is proportional to the density. This behaviour was explained in terms of the geometric deformation undergone by 
the radial metric component:  the higher the density, the more geometric deformation is produced inside the stellar 
distribution, and in consequence the induced anisotropy is higher for more compact distributions. On the other hand, it was 
found that the Weyl scalar function ${\cal U}$ is always negative inside the 
stellar distribution. This behaviour means that the geometric 
deformation producing anisotropic effects, which is a positive source for ${\cal U}$,  is not high enough to dominate the 
negative source of ${\cal U}$ produced by high energy terms inside the distribution. However this behaviour may change 
close to the surface if the compactness of the stellar distribution is near to the maximum value permitted by general relativity.

It was also shown that there are no bulk effects on the 
pressure or density in this solution. However, fixing a free parameter ($\delta\,A=0$) makes it possible to obtain the bulk 
effects $(\delta\rho, \delta\,p)$ on the density and pressure, showing that the pressure is increased by five dimensional 
effects when non-local terms are considered in uniform distributions, 
which represents a different result from other braneworld 
solution \cite{germ}, where only high energy modifications were considered, 
thus showing that the bulk Weyl curvature contributions have important consequences.

In this work the minimal geometric deformation was used in the context of the Randall-Sundrum theory with $Z_2$ symmetry. This approach might be extended in the case of braneworld theories without $Z_2$ symmetry or any junction conditions, as those introduced in \cite{maia2004} and \cite{maia2005}, which have been successfully used in the astrophysics context \cite{sepangi2009}. Another subjects of interest is the use of this approach in brane theories with variable tension, as introduced by \cite{gergely2009} in the cosmological context, and the study of codimension-2 braneworld theories, as those developed in \cite{minas2008} and \cite{papa2008}. A possible extension of this approach in all these theories is currently being investigated.

\begin{figure}
  \includegraphics{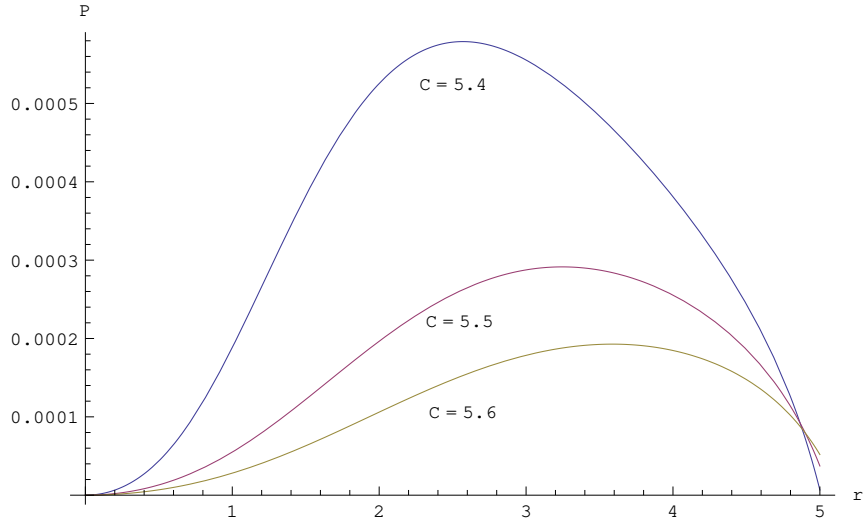}
\caption{Behaviour of the anisotropy ${\cal P}(r)$ inside the
stellar distribution with $R=5$. }
\label{fig:1}       
\end{figure}

\begin{figure}
  \includegraphics{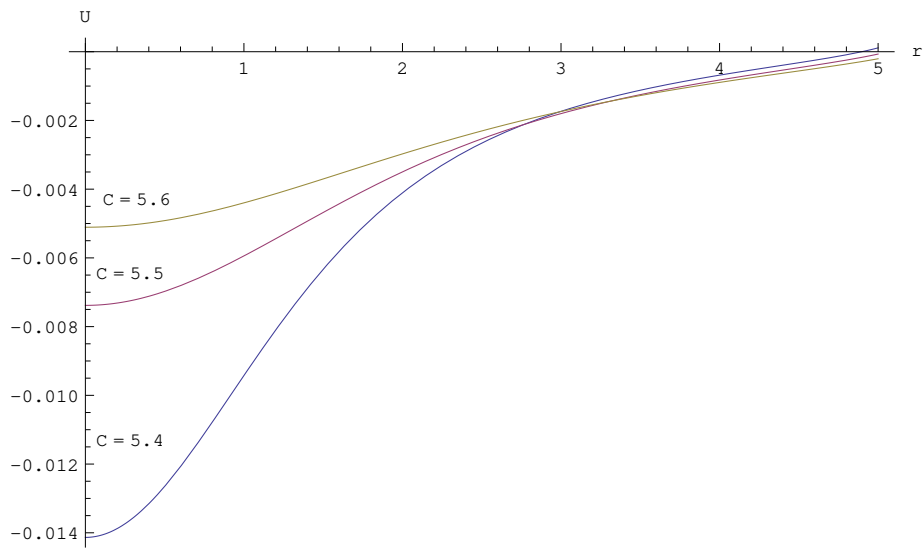}
\caption{Behaviour of the scalar Weyl function ${\cal U}(r)$ inside the
stellar distribution with $R=5$. }
\label{fig:1}       
\end{figure}

\begin{figure}
  \includegraphics{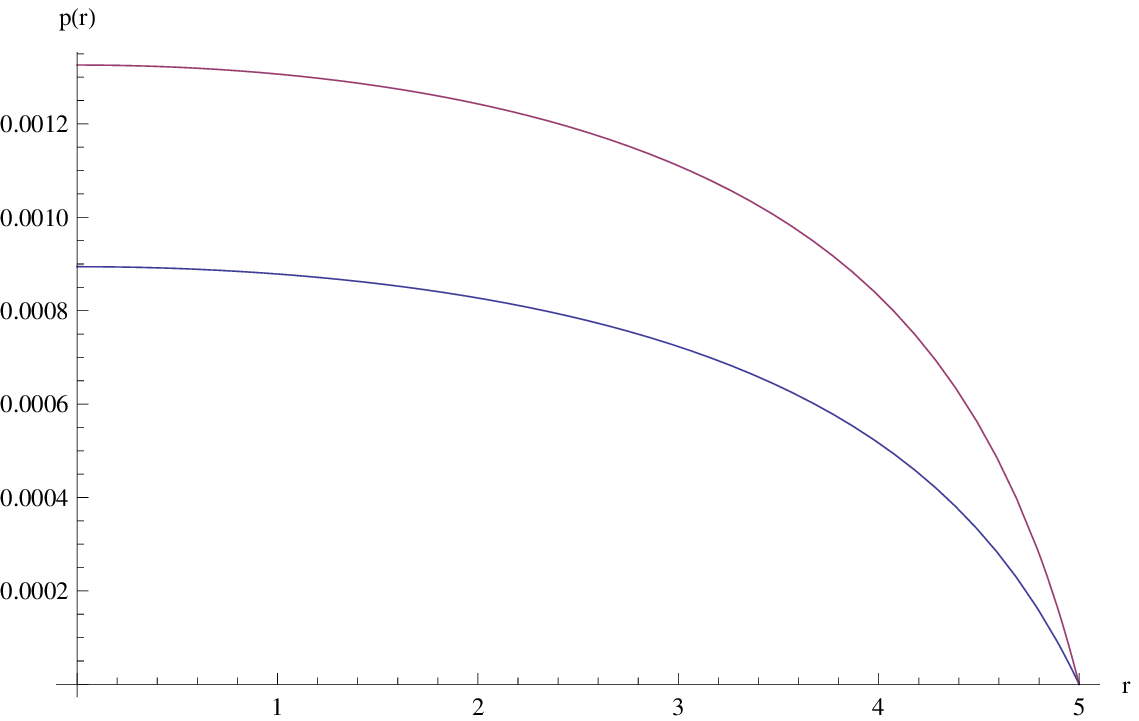}
\caption{Qualitative comparison of the pressure $p(r)$, in general relativity (lower curve) and in the braneworld model (upper curve) with $\delta\,A=0$. }
\label{fig:1}       
\end{figure}

\section*{Acknowledgments}

This work was supported by {\bf Decanato de Investigaci\'on y Desarrollo, USB}. Grant:
S1-IN-CB-002-09, and by {\bf FONACIT}. Grant: S2-2009000298.

\end{document}